\documentclass[aps,prd,showpacs,onecolumn,preprintnumbers,notitlepage,
nofootinbib,amsfont,amssymb,10pt,singlespacing] {revtex4-1}
\usepackage{amsmath,bm}
\usepackage{times}
\usepackage{braket}
\usepackage{color,graphicx}
\usepackage{slashed}
\usepackage{CJK,upgreek,fancyhdr}
\usepackage{hyperref}

\newcommand{\beq}{\begin{eqnarray}}
\newcommand{\eeq}{\end{eqnarray}}
\newcommand{\non}{\nonumber\\ }

\def\lsim{ {\ \lower-1.2pt\vbox{\hbox{\rlap{$<$}\lower6pt\vbox{\hbox{$\sim$}
}}}\ } }
\def\gsim{ {\ \lower-1.2pt\vbox{\hbox{\rlap{$>$}\lower6pt\vbox{\hbox{$\sim$}
}}}\ } }

\def \npb{  Nucl. Phys. B }

\def \plb{  Phys. Lett. B }

\def \prd{  Phys. Rev. D }

\def \jhep{ J. High Energy Phys.  }

\definecolor{Red}{rgb}{1.,0.,0.}

\definecolor{Blue}{rgb}{0.,0.,1.}

\newcommand{\orcid}[1]{\thanks{\href{http://orcid.org/#1}{ORCID: #1}}}

\definecolor{nicered}{rgb}{0.7,0.1,0.2}
\definecolor{nicegreen}{rgb}{0.1,0.4,0.2}

\hypersetup{colorlinks,citecolor=nicegreen,linkcolor=nicered}

\begin{document}
\begin{CJK*}{GB}{gbsn}
\title{ Pure annihilation decays of
\boldmath{$B_s^0 \to a_0^+ a_0^-$} and
\boldmath{$B_d^0 \to K_0^{*+} K_0^{*-}$}
in the PQCD approach }
%%%==================================================================
\author{CHEN~Yutong(³ÂÓíͬ)}
\author{JIANG~Zewen(½¯ÔóÎÄ)}
\author{LIU~Xin(ÁõÐÂ)}
\email
[ Electronic address: ]
{liuxin@jsnu.edu.cn}
\orcid{ 0000-0001-9419-7462}
\affiliation{Department of Physics,
Jiangsu Normal University, Xuzhou 221116, China}

%%%%%%%%%%%%%%%%%%%%%%%%%%%%%%%%%%%%%%%%%%%%%%%%%%%%%%%%%%%%%%%%%%%%%

\date{\today{}}

%%%%%%%%%%%%%%%%%%%%%%%%%%%%%%%%%%%%%%%%%%%%%%%%%%%%%%%%%%%%%%%%%%
\begin{abstract}
%\noindent

We study the {\it CP}-averaged branching fractions and the
{\it CP}-violating asymmetries in the pure annihilation decays of $B_s^0
\to a_0^+ a_0^-$ and $B_d^0 \to K_0^{*+} K_0^{*-}$, where $a_0\;[K_0^*]$
denotes the scalar $a_0(980)$ and $a_0(1450)$ [$K_0^*(800) ( {\rm or}\; \kappa)$
and $K_0^*(1430)$], with the perturbative QCD factorization approach
under the assumption of two-quark structure for the
$a_0$ and $K_0^*$ states. The numerical results
show that the branching ratios of the
$B_d^0 \to K_0^{*+} K_0^{*-}$ decays are in the order of $10^{-6}$, while the
decay rates of the
$B_s^0 \to a_0^+ a_0^-$ modes are in the
order of $10^{-5}$. In light of the measured modes with the same quark
components in the pseudoscalar sector, namely, $B_d^0 \to K^+ K^-$ and
$B_s^0 \to \pi^+ \pi^-$, the predictions for the considered decay modes
in this work are expected to be measured at the Large Hadron Collider
beauty and/or Belle-II experiments in the (near) future.
Meanwhile, it is of great interest to find that the twist-3 distribution
amplitudes $\phi^{S}$ and $\phi^{T}$ with inclusion of the Gegenbauer polynomials
for the scalar $a_0(1450)$ and $K_0^*(1430)$ states in scenario 2 contribute
slightly to the branching ratios while significantly to the {\it CP} violations
in the $B_d^0 \to {K_0^*}(1430)^+ {K_0^*}(1430)^-$ and $B_s^0 \to a_0(1450)^+
a_0(1450)^-$ decays, which indicates that, compared to the asymptotic $\phi^S$
and $\phi^T$, these Gegenbauer polynomials could change the strong phases
evidently in these pure annihilation decay channels.
These predictions await for the future confirmation experimentally, which could
further provide useful information to help explore the inner structure of the
scalars and shed light on the annihilation decay mechanism.

\end{abstract}

%%%%%%%%%%%%%%%%%%%%%%%%%%%%%%%%%%%%%%%%%%%%%%%%%%%%%%%%%%%%%%

\pacs{13.25.Hw, 12.38.Bx, 14.40.Nd}
\preprint{\footnotesize JSNU-PHY-HEP-01/21}
\maketitle

In the heavy flavor $B$ physics, the annihilation
diagrams are highly important
in understanding the (non-)perturbative
dynamics involved in the related decays, although which are
generally considered as the power suppressed ones
and then ever neglected because of not knowing how to
effectively calculate them at the early stage of investigating the
$B$ meson decays~\cite{Bauer:1986bm,Ali:1998gb,Beneke:1999br}~\footnote{
It is worth stressing that the annihilation
diagrams seem much more important in understanding the
dynamics contained in the charmed meson decays, e.g., see
Refs.~\cite{Li:2019hho,Yu:2017oky,Li:2013xsa,Li:2012cfa} for detail.}.
However, to interpret well the fundamental quantity, namely, the
$CP$-violating asymmetry, one realized that the annihilation diagrams
should be included essentially. Although, to date, both of the soft-collinear effective
theory~\cite{Bauer04:scet} and the pertubative QCD factorization(PQCD)
approach~\cite{Keum:2000wi,Lu:2000em,Lu:2000hj} believed that
the annihilation diagrams could be perturbatively calculated, the extremely
different observations still made this issue controversial, namely,
an almost real amplitude with a tiny strong phase was obtained
in the former framework by introducing the zero-bin subtraction~\cite{Arnesen08:anni-scet};
and in contrast, an almost imaginary one with a large strong phase appeared
in the latter formalism naturally by keeping the transverse momentum $k_T$ of
valence quark~\cite{Chay08:complexanni}. Furthermore, it is stressed that
more recent works based
on the QCD factorization approach~\cite{Beneke:1999br,Du02:qcdf},
one of the popular factorization methods
in the current market, claimed that a complex contribution arising
from the annihilation diagrams
with significant imaginary parts should be essential in the $B_{(s)} \to
PP, PV$ and $VV$ decays by fitting to experimental
data~\cite{Zhu:2011mm,Wang:2013fya,Chang:2012xv,Chang:2014rla,Chang:2014yma,
Sun:2014tfa,Chang:2015wba,Chang:2016qyc,Chang:2017brr}.
Phenomenologically speaking, they
supported the viewpoint of the PQCD approach on the effective
calculations of the annihilation diagrams
to some extent. And what is more,
the measurements from the Large Hadron Collider beauty(LHCb) experiment
on the pure annihilation modes~\cite{Aaij:2012as,Aaij:2016elb,Zyla:2020},
i.e., $B_d^0 \to K^+ K^-$ and $B_s^0 \to
\pi^+ \pi^-$, confirmed the predictions of their branching ratios
in the PQCD approach at leading order~\cite{Li:2004ep,Xiao:2011tx}.
Undoubtedly, the good agreement
between theory and experiment is very exciting and inspiring. Therefore,
in order to provide solid foundation to understand
the annihilation decay mechanism,
more and more investigations on the annihilation diagrams in the
PQCD approach are necessary.

Motivated by this success, we shall study the $B_s^0 \to a_0^+ a_0^-$
and $B_d^0 \to K_0^{*+} K_0^{*-}$ decays within the PQCD approach
at leading order in this work, where $a_0$ and $K_0^*$ denote
the light scalar states $a_0(980)$ and $a_0(1450)$, and $K_0^*(800)({\rm or}
\; \kappa)$ and $K_0^*(1430)$, respectively. Before proceeding, it
is essential to give a short review on the current status about the light scalar states
$a_0$ and $K_0^*$. It is well known that the description on the inner structure
of light scalars is still in controversial (for a review,
see e.g., Refs.~\cite{Godfrey:1998pd,Close:2002zu,Zyla:2020-scalar}).
The states $a_0(980)$ and $\kappa$
with masses below or close to 1~GeV are classified into one nonet, while those ones
$a_0(1450)$ and $K_0^*(1430)$ with masses above 1~GeV belong to the other nonet.
It has been stressed that these two nonets are hard to be considered as the
low-lying $q\bar q$ states simultaneously due to the major difficulties, e.g.,
see Ref.~\cite{Cheng:2015iaa} for detail. Therefore, two typical scenarios are suggested for the
classification of these light scalar states~\cite{Cheng:2005nb}: In scenario 1(S1), $a_0(980)$
and $\kappa$ are the lowest-lying $q\bar q$ states, while
$a_0(1450)$ and $K_0^*(1430)$ are the first excited $q\bar q$ states correspondingly;
In scenario 2(S2),  $a_0(980)$ and $\kappa$
are treated as four-quark states, and then $a_0(1450)$ and
$K_0^*(1430)$ are considered as the lowest-lying two-quark states. In the present work,
we shall consider the decays of  $B_s^0 \to a_0^+ a_0^-$
and $B_d^0 \to K_0^{*+} K_0^{*-}$ with the scalars in the two-quark model.
Then, it is easily found that the considered decays have the same quark components
as those measured ones, i.e., $B_s^0 \to \pi^+ \pi^-$ and $B_d^0 \to K^+ K^-$
in the pseudoscalar sector.
Therefore, the LHCb and/or Belle-II experiments could potentially make a sound examination
on the predictions about the branching ratios and/or the {\it CP} violations of the considered
$B_s^0 \to a_0^+ a_0^-$ and $B_d^0 \to K_0^{*+} K_0^{*-}$ decays in the PQCD approach.

At the quark level, the considered $B_s^0 \to a_0^+ a_0^-$ and $B_d^0 \to K_0^{*+} K_0^{*-}$
decays are induced by the $\bar b \to \bar s$
and the $\bar b \to \bar d$ transitions, respectively.
The related weak effective Hamiltonian $H_{\rm eff}$
can be written as~\cite{Buchalla:1995vs},
\begin{equation}
H_{\rm eff}\, =\, {G_F\over\sqrt{2}}
\left\{V_{ub}^*V_{uQ} \left[C_1(\mu)O_1^{u}(\mu)
+C_2(\mu)O_2^{u}(\mu)\right]- V_{tb}^*V_{tQ} \sum_{i=3}^{10}C_i(\mu)O_i(\mu)\right\}\;,
\label{eq:heff}
\end{equation}
with the Fermi constant $G_F=1.16639\times 10^{-5}\; {\rm GeV}^{-2}$,
the Cabibbo-Kobayashi-Maskawa(CKM) matrix elements $V_{u(t)b}$ and $V_{u(t)Q}$,
the light $Q = d, s$ quark, and the Wilson
coefficients $C_i(\mu)$ at the renormalization scale
$\mu$. The local four-quark operators $O_i(i=1,\cdots,10)$ are written as
\begin{itemize}
\item{ Tree
%current-current(tree)
operators}
\begin{eqnarray}
{\renewcommand\arraystretch{1.5}
\begin{array}{ll}
\displaystyle
O_1^{u}\, =\,
(\bar{Q}_\alpha u_\beta)_{V-A}(\bar{u}_\beta b_\alpha)_{V-A}\;,
& \displaystyle
O_2^{u}\, =\, (\bar{Q}_\alpha u_\alpha)_{V-A}(\bar{u}_\beta b_\beta)_{V-A}\;,
\end{array}}
\label{eq:operators-1}
\end{eqnarray}

\item{ QCD penguin operators}
\begin{eqnarray}
{\renewcommand\arraystretch{1.5}
\begin{array}{ll}
\displaystyle
O_3\, =\, (\bar{Q}_\alpha b_\alpha)_{V-A}\sum_{q'}(\bar{q}'_\beta q'_\beta)_{V-A}\;,
& \displaystyle
O_4\, =\, (\bar{Q}_\alpha b_\beta)_{V-A}\sum_{q'}(\bar{q}'_\beta q'_\alpha)_{V-A}\;,
\\
\displaystyle
O_5\, =\, (\bar{Q}_\alpha b_\alpha)_{V-A}\sum_{q'}(\bar{q}'_\beta q'_\beta)_{V+A}\;,
& \displaystyle
O_6\, =\, (\bar{Q}_\alpha b_\beta)_{V-A}\sum_{q'}(\bar{q}'_\beta q'_\alpha)_{V+A}\;,
\end{array}}
\label{eq:operators-2}
\end{eqnarray}

\item{ Electroweak penguin operators}
\begin{eqnarray}
{\renewcommand\arraystretch{1.5}
\begin{array}{ll}
\displaystyle
O_7\, =\,
\frac{3}{2}(\bar{Q}_\alpha b_\alpha)_{V-A}\sum_{q'}e_{q'}(\bar{q}'_\beta q'_\beta)_{V+A}\;,
& \displaystyle
O_8\, =\,
\frac{3}{2}(\bar{Q}_\alpha b_\beta)_{V-A}\sum_{q'}e_{q'}(\bar{q}'_\beta q'_\alpha)_{V+A}\;,
\\
\displaystyle
O_9\, =\,
\frac{3}{2}(\bar{Q}_\alpha b_\alpha)_{V-A}\sum_{q'}e_{q'}(\bar{q}'_\beta q'_\beta)_{V-A}\;,
& \displaystyle
O_{10}\, =\,
\frac{3}{2}(\bar{Q}_\alpha b_\beta)_{V-A}\sum_{q'}e_{q'}(\bar{q}'_\beta q'_\alpha)_{V-A}\;,
\end{array}}
\label{eq:operators-3}
\end{eqnarray}
\end{itemize}
with the color indices $\alpha, \ \beta$ and the notations
$(\bar{q}'q')_{V\pm A} = \bar q' \gamma_\mu (1\pm \gamma_5)q'$.
The index $q'$ in the summation of the above operators runs
through $u,\;d,\;s$, $c$, and $b$. Note that we will use the
leading order Wilson coefficients
since we work in the framework of the PQCD
approach at leading order.
For the renormalization group evolution of the Wilson coefficients
from higher scale to lower scale, we will adopt straightforwardly the formulas as given in
Refs.~\cite{Keum:2000wi,Lu:2000em}.

%%%%=============================================================
\begin{figure}[htb]
\centering
%\vspace{-3.0cm}
\begin{tabular}{l}
\includegraphics[width=0.5\textwidth]{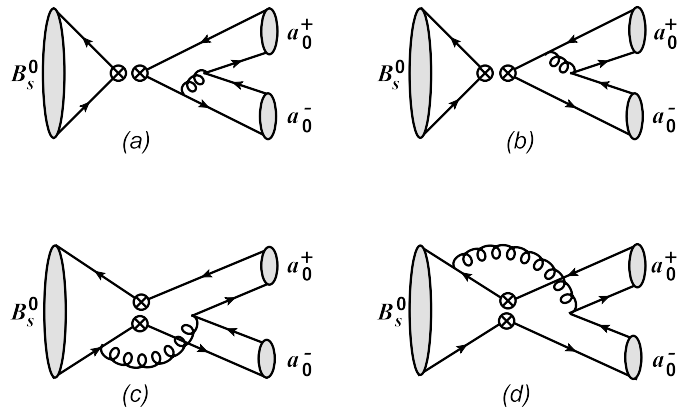}
\end{tabular}
%\vspace{-3.2cm}
\caption{Leading order Feynman diagrams for $B_s^0\to a_0^+ a_0^-$ decays in the PQCD formalism. }
\label{fig:fig1}
\end{figure}
%%%%==============================================================

The Feynman diagrams of the $B_s^0 \to a_0^+ a_0^-$
decays at leading order in the PQCD
formalism are illustrated in Fig.~\ref{fig:fig1}:
Figs.~\ref{fig:fig1}(a) and \ref{fig:fig1}(b) describe
the factorizable annihilation($fa$) diagrams,
while Figs.~\ref{fig:fig1}(c) and \ref{fig:fig1}(d)
describe the nonfactorizable annihilation($nfa$) ones.
By replacing the $s$ and $d$ quarks in the $B_s^0 \to a_0^+ a_0^-$
decays with the $d$ and $s$ ones correspondingly, then we can obtain
the annihilation modes $B_d^0 \to K_0^{*+} K_0^{*-}$ directly. As we know,
several $B \to SS$ decays with $S$ denoting the scalar mesons have been studied in the PQCD
approach~\cite{Liu:2013lka,Dou:2015mka,Su:2019vbu,Liang:2019eur,Li:2019jlp}.
Therefore, the analytic expressions
for the decay amplitudes of the considered $B_s^0 \to a_0^+ a_0^-$ and $B_d^0 \to
K_0^{*+} K_0^{*-}$ decays can be found easily, for example, in the
Refs.~\cite{Liu:2013lka,Liang:2019eur,Li:2019jlp}.
The $B_d^0 \to K_0^{*+}
K_0^{*-}$ decay amplitudes have been presented in the PQCD approach~\cite{Liu:2013lka}.
Then, we just need to
replace the $B_d^0$ and $K_0^*$ states in Ref.~\cite{Liu:2013lka} with the $B_s^0$
and $a_0$ ones, as well as the related CKM matrix elements,
to obtain easily the corresponding information of the $B_s^0 \to a_0^+ a_0^-$ decays
in the PQCD approach. Hence, for simplicity, we will not collect the aforementioned
formulas in this paper. The interested readers can refer to Ref.~\cite{Liu:2013lka}
for detail.

As presented in Ref.~\cite{Liu:2013lka}, the $B_d^0 \to K_0^{*+} K_0^{*-}$ decay amplitude
is as the following,
\beq
A(B_d^0 \to K_0^{*+} K_0^{*-}) &=& V_{ub}^* V_{ud} \biggl[ M_{nfa} C_2 \biggr] -V_{tb}^* V_{td}
\biggl\{  M_{nfa} (C_4 + C_{10}) + M_{nfa}^{P_2} (C_6 + C_8) \non
&& + M_{nfa}[K_0^{*+} \leftrightarrow K_0^{*-}] (C_4 - \frac{1}{2} C_{10})
+ M_{nfa}^{P_2}[K_0^{*+} \leftrightarrow K_0^{*-}] (C_6 - \frac{1}{2} C_8)\biggr\}\;,
\label{eq:decamp-d}
\eeq
where $M_{nfa}$ and $M_{nfa}^{P_2}$ stand for the nonfactorizable annihilation amplitudes
arising from the $(V-A)(V-A)$ and $(S-P)(S+P)$ currents~\cite{Liu:2013lka}, respectively.
Moreover, the almost exact cancellation of the factorizable annihilation amplitudes appear
in the considered two decay modes
due to the very small flavor symmetry breaking effects, which can be seen in the
numerical calculations later.
Similarly, we can easily obtain the $B_s^0 \to a_0^+ a_0^-$ decay amplitude by the
corresponding  replacement of $d \leftrightarrow s$ in the $B_d^0 \to K_0^{*+} K_0^{*-}$
one, that is,
\beq
A(B_s^0 \to a_0^{+} a_0^{-}) &=& V_{ub}^* V_{us} \biggl[ M_{nfa} C_2 \biggr] -V_{tb}^* V_{ts}
\biggl\{  M_{nfa} (C_4 + C_{10}) + M_{nfa}^{P_2} (C_6 + C_8) \non
&& + M_{nfa}[a_0^{+} \leftrightarrow a_0^{-}] (C_4 - \frac{1}{2} C_{10})
+ M_{nfa}^{P_2}[a_0^{+} \leftrightarrow a_0^{-}] (C_6 - \frac{1}{2} C_8)\biggr\}\;.
\label{eq:decamp-s}
\eeq

Then, we can turn to the numerical calculations of the {\it CP}-averaged
branching ratios and the {\it CP}-violating asymmetries of the $B_s^0 \to a_0^+ a_0^-$
and $B_d^0 \to K_0^{*+} K_0^{*-}$ decays in the PQCD approach.
Some comments on the nonperturbative inputs are listed essentially as follows:
\begin{enumerate}
\item[]{(a)} For the heavy $B_d^0$ and $B_s^0$ mesons, the wave functions and the distribution
amplitudes, and the decay constants are same as those utilized in Ref.~\cite{Liu:2013lka},
but with the updated lifetimes $\tau_{B_d^0} = 1.52$~ps and $\tau_{B_s^0} = 1.509$~ps
~\cite{Tanabashi:2018oca}. {\it It is worth mentioning that, due to its highly small effects, namely, the power-suppressed $1/m_B$ contributions to $B$ decays in final states with
energetic light
particles~\cite{Lu:2000em,Braun:2017liq}, the high twist contributions from the $B$ meson wave function in the considered pure annihilation channels have to be left for future studies associated with the precise measurements}.
For recent development about the $B$ meson wave function and/or
distribution amplitude, please see references, e.g.,~\cite{Braun:2017liq,Galda:2020epp,Wang:2019msf,Li:2014xda,Feldmann:2014ika,Bell:2013tfa} for detail.

\item[]{(b)}
For the considered light scalar $a_0$ and $K_0^*$ states, the decay constants and
the Gegenbauer moments in the distribution amplitudes~\footnote{It
is necessary to mention that we firstly adopt the asymptotic form of the twist-3
distribution amplitudes $\phi^S$ and $\phi^T$(T3A) in the numerical calculations
here as usual~\cite{Cheng:2005nb,Li:2008tk}. And then we will estimate the
effects in this work arising from the
twist-3 distribution amplitudes with inclusion of the Gegenbauer polynomials(T3G)
in S2 later. It is noted that only the T3G form in S2
is available currently~\cite{Lu:2006fr}.} have
been derived at the normalization scale $\mu=1$ GeV in the QCD sum rule
method~\cite{Cheng:2005nb}:
\beq
    \bar f_{\kappa} &=& 0.340 \pm 0.020 \;\; {\rm GeV}\;, \;\; f_{\kappa} = 0.050 \pm 0.003 \;\; {\rm GeV}\;,
    %\;\; ({\rm S1})
    \label{eq:decs-ka}\\
    B_1 &=& -0.92 \pm 0.11 \;, \;\;\;\hspace{0.81cm} B_3= 0.15 \pm 0.09\;;
    %\hspace{1.450cm}({\rm S1})
    \label{eq:gb-ka}
\eeq
\beq
    \bar f_{a_0(980)} &=& 0.365 \pm 0.020 \;\; {\rm GeV}\;, \;\; f_{a_0(980)} \sim 0.0011 \;\; {\rm GeV}\;,
    %\;\; ({\rm S1})
    \\
    B_1 &=& -0.93 \pm 0.10 \;, \;\;\;\hspace{1.58cm} B_3 = 0.14 \pm 0.08\;;
    %\hspace{0.40cm}({\rm S1})
\eeq
\beq
    \bar f_{K_0^*(1430)} &=& \left\{ \begin{array}{ll}
-0.300 \pm 0.030 \;\; {\rm GeV}\;,  &  \\
\hspace{0.28cm}0.445 \pm 0.050 \;\; {\rm GeV} \;,  &   \\ \end{array} \right.
     f_{K_0^*(1430)} =
\left\{ \begin{array}{ll}
-0.044^{+0.004}_{-0.005} \;\; {\rm GeV}\;,& \;\; ({\rm S1})  \\
\hspace{0.28cm}0.066^{+0.007}_{-0.008} \;\; {\rm GeV}\;;  & \;\; ({\rm S2})   \\ \end{array} \right. \label{eq:decs-k0s}\\
    B_1 &=&
\left\{ \begin{array}{ll}
\hspace{0.28cm}0.58 \pm 0.07\;,  &  \\
-0.57 \pm 0.13\;,  &   \\ \end{array} \right.
\;\;\hspace{2.01cm}    B_3 =
\left\{ \begin{array}{ll}
-1.20 \pm 0.08\;,& \;\; ({\rm S1})  \\
-0.42 \pm 0.22\;; & \;\; ({\rm S2})   \\ \end{array} \right.
\label{eq:gb-k0s}
\eeq
\beq
    \bar f_{a_0(1450)} &=&
\left\{ \begin{array}{ll}
-0.280 \pm 0.030 \;\; {\rm GeV}\;,  &  \\
\hspace{0.28cm}0.460 \pm 0.050 \;\; {\rm GeV}\;,   &   \\ \end{array} \right.
     f_{a_0(1450)} =
\left\{ \begin{array}{ll}
\sim -0.0009 \hspace{0.37cm}  \;\; {\rm GeV}\;,&  \;\; ({\rm S1}) \\
0.0014^{+0.0002}_{-0.0001} \;\; {\rm GeV}\;; & \;\; ({\rm S2})   \\ \end{array} \right.
\label{eq:decs-a14}\\
    B_1 &=&
\left\{ \begin{array}{ll}
\hspace{0.28cm}0.89 \pm 0.20\;,  &  \\
-0.58 \pm 0.12\;,  &   \\ \end{array} \right.
\;\;\hspace{1.92cm}    B_3 =
\left\{ \begin{array}{ll}
-1.38 \pm 0.18\;,& \;\; ({\rm S1})  \\
-0.49 \pm 0.15\;. & \;\; ({\rm S2})   \\ \end{array} \right.
\label{eq:gb-a14}
\eeq
Note that the scale-dependent scalar decay constant $\bar f_{S}$ and
the vector decay constant $f_{S}$ are related
with each other through the following relation~\cite{Cheng:2005nb},
\beq
\bar f_{S} &=& \mu_S f_{S} = \frac{m_S}{m_{q_2}(\mu) - m_{q_1}(\mu)} f_S \;,
\label{eq:s-v-DC}
\eeq
in which $m_S$ is the mass of the scalar meson, and $m_{q_2}$ and $m_{q_1}$ are
the running quark masses. It is worth pointing out that for the $a_0$ states,
the isospin symmetry breaking effects from the $u$ and $d$ quark masses are considered.
Therefore, the running quark masses for the strange quark and the nonstrange
light quarks can be read as $m_s = 0.128$~GeV, $m_d = 0.006$~GeV, and $m_u = 0.003$~GeV,
respectively, which are translated from those
at the $\overline{\rm MS}$ scale $\mu \approx 2$ GeV~\cite{Tanabashi:2018oca}.
For the masses of the $a_0$ and $K_0^*$ states, the values $m_{\kappa} = 0.824$~GeV,
$m_{a_0(980)} = 0.980$~GeV, $m_{K_0^*(1430)} = 1.425$~GeV, and $m_{a_0(1450)} = 1.474$~GeV
will be adopted in the numerical calculations
~\footnote{ As inferred from the {\it Review of Particle Physics}~\cite{Zyla:2020},
the considered scalar $a_0$ and $K_0^*$ states are also with finite
widths. Generally speaking, the width effect could change the numerical results
with different extent~\cite{Cheng:2003xc,Cheng:2020mna}. In principle, we should
consider the width effect to make relevant predictions more precise. However,
it is unfortunate that the distribution amplitudes for the considered $S$-wave
resonance states with the constrained parameters, e.g., Gegenbauer moments,
are currently unavailable. Therefore, we have to leave the width effect in
this work for future investigations elsewhere.}.

\item[]{(c)} For the CKM matrix elements, we also adopt the Wolfenstein
parametrization at leading order, but with the updated parameters $A=0.836$,
$\lambda=0.22453$, $\bar \rho= 0.122^{+0.018}_{-0.017}$, and $\bar \eta=
0.355^{+0.012}_{-0.011}$~\cite{Tanabashi:2018oca}.
\end{enumerate}

Now, we present the numerical results of the $B_s^0 \to a_0^+ a_0^-$ and $B_d^0
\to K_0^{*+} K_0^{*-}$ decays in the PQCD formalism. Firstly, the PQCD predictions
of the {\it CP}-averaged branching ratios
can be read as follows:
\beq
Br(B_d^0 \to \kappa^+ \kappa^-) &=&
%2.84^{+0.19}_{-0.21}(\omega_B)^{+1.36+0.41}_{-0.99-0.30}(B_{K_0^*})^{+0.74}_{-0.61}
%(\bar f_{K^*_0})^{+0.15+0.09}_{-0.13-0.07}(V)
2.84^{+0.19}_{-0.21}(\omega_B)^{+1.42}_{-1.03}(B_m)^{+0.74}_{-0.61}(\bar f_{K^*_0})^{+0.17}_{-0.15}(V)[2.84^{+1.62}_{-1.22}]
\times 10^{-6}\;,
%\;\; ({\rm S1})
%\non
%&=&2.84^{+1.62}_{-1.22} \times 10^{-6}\;,
\label{eq:br-kappa}\\
%\eeq
%%
%\beq
Br(B_s^0 \to a_0(980)^+ a_0(980)^-) &=&
%2.64^{+0.30}_{-0.28}(\omega_B)^{+1.19+0.07}_{-0.88-0.00}(B_{a_0})^{+0.63}_{-0.53}
%(\bar f_{a_0})^{+0.02+0.00}_{-0.02-0.00}(V)
2.64^{+0.30}_{-0.28}(\omega_B)^{+1.19}_{-0.88}(B_m)^{+0.63}_{-0.53}
(\bar f_{a_0})^{+0.02}_{-0.02}(V)[2.64^{+1.38}_{-1.06}]
\times 10^{-5}\;,
%\;\; ({\rm S1})
%\non
%&=&2.64^{+1.38}_{-1.06} \times 10^{-5}\;,
\label{eq:br-a0}
\eeq
and
\beq
Br(B_d^0 \to {K_0^{*}}(1430)^+ {K_0^{*}}(1430)^-) &=&
\left\{ \begin{array}{ll}
%3.08^{+0.00}_{-0.07}(\omega_B)^{+0.66+0.52}_{-0.60-0.48}(B_{K_0^*})^{+1.42}_{-1.06}
%(\bar f_{K_0^*})^{+0.11+0.11}_{-0.10-0.11}(V)
3.08^{+0.00}_{-0.07}(\omega_B)^{+0.84}_{-0.77}(B_m)^{+1.42}_{-1.06}
(\bar f_{K_0^*})^{+0.16}_{-0.15}(V)[3.08^{+1.66}_{-1.32}]
\times 10^{-6}\;,&\;\; ({\rm S1})  \\
%2.45^{+0.51}_{-0.43}(\omega_B)^{+1.64+1.07}_{-1.08-0.61}(B_{K_0^*})^{+1.58}_{-0.93}
%(\bar f_{K_0^*})^{+0.09+0.03}_{-0.09-0.03}(V)
2.45^{+0.51}_{-0.43}(\omega_B)^{+1.96}_{-1.24}(B_m)^{+1.58}_{-0.93}
(\bar f_{K_0^*})^{+0.09}_{-0.09}(V)[2.45^{+2.57}_{-1.61}]
\times 10^{-6}\;; &\;\; ({\rm S2})   \\ \end{array} \right.
\label{eq:br-k1430}\\
%\eeq
%%
%\beq
Br(B_s^0 \to a_0(1450)^+ a_0(1450)^-) &=&
\left\{ \begin{array}{ll}
%5.63^{+0.00}_{-0.16}(\omega_B)^{+2.35+2.10}_{-1.78-1.62}(B_{a_0})^{+2.83}_{-2.05}
%(\bar f_{a_0})^{+0.04+0.00}_{-0.04-0.00}(V)
5.63^{+0.00}_{-0.16}(\omega_B)^{+3.15}_{-2.41}(B_m)^{+2.83}_{-2.05}
(\bar f_{a_0})^{+0.04}_{-0.04}(V)[5.63^{+4.23}_{-3.17}]
\times 10^{-5}\;,&\;\; ({\rm S1})  \\
%3.55^{+0.96}_{-0.75}(\omega_B)^{+2.15+0.89}_{-1.48-0.64}(B_{a_0})^{+1.82}_{-1.31}
%(\bar f_{a_0})^{+0.01+0.00}_{-0.01-0.00}(V)
3.55^{+0.96}_{-0.75}(\omega_B)^{+2.33}_{-1.61}(B_m)^{+1.82}_{-1.31}
(\bar f_{a_0})^{+0.01}_{-0.01}(V)[3.55^{+3.11}_{-2.21}]
\times 10^{-5}\;, &\;\; ({\rm S2})   \\ \end{array} \right.
\label{eq:br-a1450}
\eeq
where, as clearly seen from the above results, the majored errors are mainly induced
by the uncertainties of the scalar decay constants $\bar f_{a_0}$ and $\bar f_{K_0^*}$,
and the combined Gegenbauer moments $B_m$ of $B_1$ and $B_3$ in the leading
twist distribution amplitudes of
the scalar mesons. The other errors induced by the shape parameter $\omega_B$
in the $B_{d,s}^0$
meson distribution amplitude and by the combined CKM matrix elements $V$ are much smaller.
Frankly speaking, these mentioned hadronic parameters of the scalar mesons are currently
less constrained from the experiments and/or Lattice QCD calculations. Therefore, we have to
adopt those available parameters calculated in the QCD sum rule method to give a rough
estimation preliminarily. Of course, both the essential measurements at the experimental
aspects and the Lattice QCD computation at the theoretical aspects on the
above-mentioned nonperturbative
inputs for the scalar mesons are urgently demanded, which is expected to help
better understand the related hadron dynamics and provide more precise predictions.
Note that all the errors from various parameters as specified above have been
added in quadrature, which can be seen from the results presented in the square brackets.

Based on the numerical results with large theoretical uncertainties as shown in
the Eqs.~(\ref{eq:br-kappa})-(\ref{eq:br-a1450}), several remarks are in order:
\begin{itemize}
\item[]{(a)}
For the considered $B_d^0 \to K_0^{*+} K_0^{*-}$ decays,
the {\it CP}-averaged branching ratios presented in Refs.~\cite{Liu:2013lka,Su:2019vbu}
could be reproduced in this work but with slight differences, which are mainly because
of the updated parameters such as the values of the CKM matrix elements,
the $B_d^0$ meson lifetime, and the running quark masses of the light quarks.
As aforementioned, the $B_d^0 \to K^+ K^-$ decay has been measured experimentally
and the decay rate is read as $7.8 \pm 1.5 \times 10^{-8}$~\cite{Zyla:2020},
which confirmed the PQCD calculation, that is, $1.56 ^{+0.56}_{-0.52}\times 10^{-7}$~\cite{Xiao:2011tx}
about this channel theoretically.
As the counterpart with the same
quark structure in the scalar sector, the $B_d^0 \to K_0^{*+} K_0^{*-}$ modes with large
decay rates in the order of $10^{-6}$ predicted with the same PQCD formalism could be
examined at the LHCb and/or Belle-II experiments in the (near) future. The confirmation
from the experimental side would help explore the inner structure especially for the scalar
$K_0^*(1430)$ meson.

\item[]{(b)}
For the $B_s^0 \to a_0^+ a_0^-$ modes, they have the same quark
structure as that of the measured one $B_s^0 \to \pi^+ \pi^-$,
whose decay rate is read as $7.0 \pm 1.0 \times 10^{-7}$~\cite{Zyla:2020}
and agrees well with the prediction about the decay rate $5.10^{+2.26}_{-1.89}
\times 10^{-7}$~\cite{Xiao:2011tx} in the PQCD approach
within theoretical errors.
Therefore, the large $B_s^0 \to a_0^+ a_0^-$
decay rates in the order of $10^{-5}$ could be accessed with much more probabilities
experimentally with respect to those of the $B_d^0 \to K_0^{*+} K_0^{*-}$ ones.
Furthermore,
\begin{itemize}
\item[]{(i)}
Recently, the $B_s^0 \to a_0(980)^+ a_0(980)^-$ mode has been investigated in
the PQCD approach by Liang and Yu~\cite{Liang:2019eur}. However, it is a bit
strange that
the {\it CP}-averaged branching ratio of the $B_s^0 \to  a_0(980)^+ a_0(980)^-$ channel
is predicted as small as $5.17^{+2.36}_{-1.94} \times 10^{-6}$. In fact,
as naively estimated from the $B_s^0 \to \pi^+ \pi^-$ decay rate, the
$B_s^0 \to  a_0(980)^+ a_0(980)^-$
branching ratio could be, as far as the central values
are concerned, $|\frac{\bar f_{a_0(980)}}{f_\pi}|^2 \cdot
|\frac{\bar f_{a_0(980)}}{f_\pi}|^2 \cdot Br(B_s^0 \to \pi^+ \pi^-)_{\rm Exp}
\sim 4.34\times 10^{-5}$ and $|\frac{\bar f_{a_0(980)}}{f_\pi}|^2 \cdot
|\frac{\bar f_{a_0(980)}}{f_\pi}|^2 \cdot Br(B_s^0 \to \pi^+ \pi^-)_{\rm PQCD}
\sim 3.16 \times 10^{-5}$ singly induced by the decay constants
$f_{\pi} =0.13$~GeV and $\bar f_{a_0(980)} = 0.365$~GeV, apart from the
possible enhancement arising from the asymmetric behavior of the only odd
terms involved in the leading twist distribution amplitude of the scalar meson.

\item[]{(ii)}
We study the $B_s^0 \to a_0(1450)^+ a_0(1450)^-$ decay in two scenarios within the PQCD
framework for the first time. The predicted branching ratios are large around $10^{-5}$
and are expected to be tested in the near future at LHCb and/or Belle-II experiments.
The understanding of the scalar $a_0(1450)$ meson could provide useful information
to uncover the nature of the $K_0^*(1430)$ meson through the $SU(3)$ flavor
symmetry (breaking) effects, and vice versa.
\end{itemize}

\item[]{(c)}
In light of the large theoretical errors, a precise ratio of the related branching
ratios would be more interested because, generally speaking, the theoretical errors
resulted from the hadronic inputs could be cancelled to a great extent. Therefore,
we define the following ratios to be measured at the relevant experiments of $B$
meson decays, which would help to study the QCD dynamics, even the decay mechanism of
these considered pure annihilation decays.
\beq
R^{a_0(980)/\kappa}_{sd}&\equiv& \frac{Br(B_s^0 \to a_0(980)^+ a_0(980)^-)}{Br(B_d^0
\to \kappa^+ \kappa^-)} \non
&=& 9.30^{+0.40}_{-0.33}(\omega_B)^{+0.42}_{-0.31}(B_m)^{+0.16}_{-0.17}
(\bar f_M)^{+0.44}_{-0.46}(V)[9.30^{+0.75}_{-0.67}]\;,
\label{eq:R1}
\eeq
\beq
R_{sd}^{a_0(1450)/K_0^*(1430)}&\equiv& \frac{Br(B_s^0 \to a_0(1450)^+ a_0(1450)^-)}
{Br(B_d^0 \to {K_0^*}(1430)^+ {K_0^*}(1430)^-)}\non
&=&
\left\{ \begin{array}{ll}
18.28^{+0.00}_{-0.11}(\omega_B)^{+4.12}_{-4.34}(B_{m})^{+0.52}_{-0.56}
(\bar f_{M})^{+0.80}_{-0.78}(V)[18.28^{+4.23}_{-4.45}] \;,&\;\; ({\rm S1})  \\
14.49^{+0.75}_{-0.63}(\omega_B)^{+1.54}_{-0.86}(B_{m})^{+0.25}_{-1.16}
(\bar f_{M})^{+0.51}_{-0.47}(V)[14.49^{+1.80}_{-1.64}] \;. &\;\; ({\rm S2})   \\
\end{array} \right.
\label{eq:R2}
\eeq
\beq
R_d^{{\rm S1}/{\rm S2}}(K_0^*(1430))&\equiv& \frac{Br(B_d^0 \to {K_0^*}(1430)^+
{K_0^*}(1430)^-)_{\rm S1}}{Br(B_d^0 \to {K_0^*}(1430)^+ {K_0^*}(1430)^-)_{\rm S2}} \non
&=& 1.26^{+0.23}_{-0.22}(\omega_B)^{+0.65}_{-0.37}(B_{m})^{+0.07}_{-0.14}
(\bar f_{M})^{+0.02}_{-0.02}(V)[1.26^{+0.69}_{-0.45}] \;,
\label{eq:R3}
\eeq
\beq
R_s^{{\rm S1}/{\rm S2}}(a_0(1450))&\equiv& \frac{Br(B_s^0 \to a_0(1450)^+
a_0(1450)^-)_{\rm S1}}{Br(B_s^0 \to a_0(1450)^+ a_0(1450)^-)_{\rm S2}} \non
&=& 1.59^{+0.36}_{-0.34}(\omega_B)^{+0.07}_{-0.10}(B_{m})^{+0.01}_{-0.01}
(\bar f_{M})^{+0.00}_{-0.01}(V)[1.59^{+0.37}_{-0.35}] \;,
\label{eq:R4}
\eeq
\beq
R_{d, {\rm S1}}^{K_0^*(1430)/\kappa}&\equiv& \frac{Br(B_d^0 \to
{K_0^*}(1430)^+ {K_0^*}(1430)^-)_{\rm S1}}{Br(B_d^0 \to \kappa^+ \kappa^-)} \non
&=& 1.08^{+0.06}_{-0.06}(\omega_B)^{+0.20}_{-0.16}(B_{m})^{+0.18}_{-0.17}
(\bar f_{M})^{+0.01}_{-0.00}(V)[1.08^{+0.28}_{-0.24}] \;,
\label{eq:R5}
\eeq
\beq
R_{d, {\rm S2}}^{K_0^*(1430)/\kappa}&\equiv& \frac{Br(B_d^0 \to
{K_0^*}(1430)^+ {K_0^*}(1430)^-)_{\rm S2}}{Br(B_d^0 \to \kappa^+ \kappa^-)} \non
&=& 0.86^{+0.12}_{-0.09}(\omega_B)^{+0.18}_{-0.19}(B_{m})^{+0.27}_{-0.18}
(\bar f_{M})^{+0.02}_{-0.04}(V)[0.86^{+0.35}_{-0.28}] \;,
\label{eq:R6}
\eeq
\beq
R_{s, {\rm S1}}^{a_0(1450)/a_0(980)}&\equiv& \frac{Br(B_s^0 \to a_0(1450)^+
a_0(1450)^-)_{\rm S1}}{Br(B_s^0 \to a_0(980)^+ a_0(980)^-)} \non
&=& 2.13^{+0.19}_{-0.22}(\omega_B)^{+0.16}_{-0.30}(B_{m})^{+0.46}_{-0.43}
(\bar f_{M})^{+0.00}_{-0.00}(V)[2.13^{+0.52}_{-0.57}] \;,
\label{eq:R7}
\eeq
\beq
R_{s, {\rm S2}}^{a_0(1450)/a_0(980)}&\equiv& \frac{Br(B_s^0 \to a_0(1450)^+
a_0(1450)^-)_{\rm S2}}{Br(B_s^0 \to a_0(980)^+ a_0(980)^-)} \non
&=& 1.34^{+0.19}_{-0.15}(\omega_B)^{+0.20}_{-0.24}(B_{m})^{+0.30}_{-0.28}
(\bar f_{M})^{+0.01}_{-0.00}(V)[1.34^{+0.41}_{-0.40}]\;.
\label{eq:R8}
\eeq
Of course, it is found that the uncertainties in some of the above ratios are not small,
for example, $R_d^{\rm S1/S2}(K_0^*(1430))=1.26^{+0.69}_{-0.45}$ in Eq.~(\ref{eq:R3}).
The underlying reason is that the large uncertainties induced by the Gegenbauer moments
$B_1$ and $B_3$ cannot be cancelled correspondingly, unlike the exact cancellation of the
uncertainties resulted from the scalar decay constants that can be isolated from
the distribution amplitudes.

\item[]{(d)}
Another eight more interesting ratios could be obtained and are expected to be
examined at the future experiments, if we take the already measured $B_d^0 \to K^+
K^-$ and $B_s^0 \to \pi^+ \pi^-$ decays as referenced channels. By combing the
branching fractions of the $B_d^0 \to K^+ K^-$ and $B_s^0 \to \pi^+ \pi^-$ decays
from both of the PQCD predictions~\cite{Xiao:2011tx} and the experimental measurements~\cite{Zyla:2020} sides, and the decay rates
of the $B_d^0 \to K_0^{*+} K_0^{*-}$ and $B_s^0 \to a_0^+ a_0^-$ modes in this work,
they are read as follows,
\beq
R_{d,{\rm Exp}}^{\kappa/K} &\equiv&
\frac{Br(B_d^0 \to \kappa^+ \kappa^-)_{\rm PQCD}}{Br(B_d^0 \to K^+ K^-)_{\rm Exp}}
= 36.41^{+11.55}_{-10.70}  \;,
\eeq
\beq
R_{d, {\rm Exp}}^{K_0^*(1430)/K}&\equiv&
\frac{Br(B_d^0 \to {K_0^*}(1430)^+ {K_0^*}(1430)^-)_{\rm PQCD}}
{Br(B_d^0 \to K^+ K^-)_{\rm Exp}}
= \left\{ \begin{array}{ll}
39.49^{+11.48}_{-11.55} \;,&\;\; ({\rm S1})  \\
31.41^{+22.57}_{-18.08} \;; &\;\; ({\rm S2})   \\ \end{array} \right.
\eeq
\beq
R_{d,{\rm Th}}^{\kappa/K} &\equiv& \frac{Br(B_d^0 \to \kappa^+
\kappa^-)_{\rm PQCD}}{Br(B_d^0 \to K^+ K^-)_{\rm PQCD}}
=  18.21^{+2.83}_{-2.63}  \;,
\eeq
\beq
R_{d, {\rm Th}}^{K_0^*(1430)/K}&\equiv&
\frac{Br(B_d^0 \to {K_0^*}(1430)^+ {K_0^*}(1430)^-)_{\rm PQCD}}
{Br(B_d^0 \to K^+ K^-)_{\rm PQCD}}
=  \left\{ \begin{array}{ll}
19.74^{+2.62}_{-2.82} \;,&\;\; ({\rm S1})  \\
15.71^{+7.97}_{-7.63} \;; &\;\; ({\rm S2})   \\ \end{array} \right.
\eeq
and
\beq
R_{s,{\rm Exp}}^{a_0(980)/\pi} &\equiv& \frac{Br(B_s^0 \to a_0(980)^+
a_0(980)^-)_{\rm PQCD}}{Br(B_s^0 \to \pi^+ \pi^-)_{\rm Exp}}
=  37.18^{+13.07}_{-10.85}  \;,
\eeq
\beq
R_{s, {\rm Exp}}^{a_0(1450)/\pi}&\equiv&
\frac{Br(B_s^0 \to a_0(1450)^+ a_0(1450)^-)_{\rm PQCD}}
{Br(B_s^0 \to \pi^+ \pi^-)_{\rm Exp}}
=   \left\{ \begin{array}{ll}
80.43^{+42.82}_{-39.43} \;,&\;\; ({\rm S1})  \\
50.71^{+32.54}_{-28.38} \;; &\;\; ({\rm S2})   \\ \end{array} \right.
\eeq
\beq
R_{s,{\rm Th}}^{a_0(980)/\pi} &\equiv& \frac{Br(B_s^0 \to a_0(980)^+
a_0(980)^-)_{\rm PQCD}}{Br(B_s^0 \to \pi^+ \pi^-)_{\rm PQCD}}
= 51.76^{+2.86}_{-2.54}   \;,
\eeq
\beq
R_{s, {\rm Th}}^{a_0(1450)/\pi}&\equiv&
\frac{Br(B_s^0 \to a_0(1450)^+ a_0(1450)^-)_{\rm PQCD}}
{Br(B_s^0 \to \pi^+ \pi^-)_{\rm PQCD}}
=  \left\{ \begin{array}{ll}
110.39^{+23.58}_{-33.75} \;,&\;\; ({\rm S1})  \\
\hspace{0.18cm}69.61^{+20.88}_{-27.87} \;. &\;\; ({\rm S2})   \\ \end{array} \right.
\eeq
These large values of the above ratios with still large theoretical errors could be easily tested when the related samples
are collected with good precision experimentally.

\item[]{(e)}
As mentioned in the above, the isospin symmetry breaking effects from the $u$ and $d$
quark masses have been considered in the $B_s^0 \to a_0^+ a_0^-$ decays.
Therefore, the factorizable annihilation decay amplitudes are not exact zero, which can
be seen clearly from the numerical results as shown in the Tables~\ref{tab:DeAms-S1}
and~\ref{tab:DeAms-S2} in both scenarios S1 and S2. However, they are still tiny
and could be neglected safely. It means that the contributions to the pure
annihilation decays considered in this work are absolutely from the nonfactorizable
annihilation decay amplitudes. It is noticed that, relative to the $B_d^0 \to
K^+ K^-$ and $B_s^0 \to \pi^+ \pi^-$ decays, the
antisymmetric QCD behavior of the leading twist distribution amplitude
could make the destructive interferences in the pseudoscalar sector
become the constructive ones in the scalar sector to the nonfactorizable annihilation
diagrams between the Fig.~\ref{fig:fig1}(c) with hard gluon radiating from
light $d(s)$ quark and the Fig.~\ref{fig:fig1}(d) with hard gluon radiating
from heavy anti-$b$ quark, which eventually
result in the large {\it CP}-averaged decay rates of the considered
decays, as presented in the Eqs.~(\ref{eq:br-kappa})-(\ref{eq:br-a1450}).

%%%================================================================================
\begin{table}[htb]
\caption{The factorization decay amplitudes(in units of $10^{-3}$~GeV$^{3}$) in S1
of the pure annihilation $B_d^0 \to K_{0}^{*+} K_{0}^{*-}$ and
$B_s^0\to a_{0}^+ a_{0}^-$ decays in the PQCD approach,
where only the central values are
quoted for clarifications. }
\label{tab:DeAms-S1}
\begin{center}\vspace{-0.5cm}{%\footnotesize
\begin{tabular}[t]{c||c|c}
\hline  \hline
   Modes
   & $A_{nfa}$(T3A) &$A_{fa}$(T3A) \\
\hline
 $B_d^0 \to {K_{0}^{*}(800)}^+ {K_{0}^{*}(800)}^-$
     &$-0.610 - {\it i} 1.974$
     &$0.0008 + {\it i} 0.0004$
 \\
 \hline
 $B_d^0 \to {K_{0}^{*}(1430)}^+ {K_{0}^{*}(1430)}^-$
    & $1.877 - {\it i} 2.658  $
    & $0.001 + {\it i} 0.003 $
 \\
 \hline\hline
 $B_s^0 \to a_0(980)^+ a_0(980)^-$
     &$6.892 + {\it i} 1.325$
     &$0.00003 - {\it i} 0.002$
 \\
 \hline
 $B_s^0 \to a_0(1450)^+ a_0(1450)^-$
     & $-1.588 + {\it i} 10.057 $
     & $-0.006 + {\it i} 0.002 $
 \\
 \hline \hline
\end{tabular}}
\end{center}
\end{table}
%%===========================================================

%%%================================================================================
\begin{table}[htb]
\caption{
Similar to Table~\ref{tab:DeAms-S1} but in S2 for the
$B_d^0 \to {K_0^*}(1430)^+ {K_0^*}(1430)^-$ and
$B_s^0\to a_{0}(1450)^+ a_{0}(1450)^-$ decays.
}
\label{tab:DeAms-S2}
\begin{center}\vspace{-0.5cm}{%\footnotesize
\begin{tabular}[t]{c||c|c||c|c}
\hline  \hline
   Modes
   & $A_{nfa}$(T3A) &$A_{fa}$(T3A)& $A_{nfa}$(T3G) &$A_{fa}$(T3G) \\
 \hline
 $B_d^0 \to {K_{0}^{*}(1430)}^+ {K_{0}^{*}(1430)}^-$
    & $ -1.940 + {\it i} 0.188 $
    & $  -0.0005 - {\it i} 0.0006$
    & $ -1.617 + {\it i} 1.571 $
    & $  -0.028 + {\it i} 0.762$
 \\
 \hline\hline
 $B_s^0 \to a_0(1450)^+ a_0(1450)^-$
     & $ 5.038 - {\it i} 6.778$
     & $-0.002 + {\it i} 0.00006$
     & $ 4.851 - {\it i} 6.812$
     & $0.006 + {\it i} 0.003$
 \\
 \hline \hline
\end{tabular}}
\end{center}
\end{table}
%%=====================================

\end{itemize}

Next, we will discuss the {\it CP}-violating asymmetries of the $B_d^0 \to K_0^{*+} K_0^{*-}$
and $B_s^0 \to a_0^+ a_0^-$ decays in the PQCD approach. Similar to the {\it CP} violations
discussed in Ref.~\cite{Liu:2013lka}, we will present the direct
and the mixing-induced {\it CP} violations ${\cal A}_{\rm dir}$ and
${\cal A}_{\rm mix}$ for the $B_d^0 \to K_0^{*+} K_0^{*-}$ decays. While, except for
${\cal A}_{\rm dir}$ and
${\cal A}_{\rm mix}$, the third {\it CP} asymmetry ${\cal A}_{\rm \Delta\Gamma_s}$ should
be considered simultaneously for the $B_s^0 \to
a_0^+ a_0^-$ decays because of the nonzero ratio $(\Delta\Gamma/\Gamma)_{B_s^0}$
for the $B_s^0-\bar{B}_s^0$ mixing, where $\Delta\Gamma$
is the decay width difference of the
$B_s^0$ meson mass eigenstates~\cite{Beneke99:Bsmixing,Fernandez06:Bsmixing}.
Then the numerical results of the {\it CP} asymmetries ${\cal A}_{\rm dir}$,
${\cal A}_{\rm mix}$, even ${\cal A}_{\rm \Delta\Gamma_s}$~\footnote{\it The definitions of
${\cal A}_{\rm dir}$, ${\cal A}_{\rm mix}$, and ${\cal A}_{\rm \Delta\Gamma_s}$
are as follows: ${\cal A}_{\rm dir}=
\frac{|\lambda_{\rm CP}|^2-1}{|\lambda_{\rm CP}|^2+1}$, ${\cal A}_{\rm mix}= \frac{2{\rm Im}(\lambda_{\rm CP})}{|\lambda_{\rm CP}|^2+1}$, and ${\cal A}_{\rm \Delta\Gamma_s}= \frac{2{\rm Re}(\lambda_{\rm CP})}{|\lambda_{\rm CP}|^2+1}$, respectively, where the {\it CP}-violating parameter
$\lambda_{\rm CP} \equiv \eta_f \frac{V_{tb}^*V_{td,s}}{V_{tb}V_{td,s}^*}\cdot \frac{\langle f_{\rm CP}|H_{eff}|\bar B_{d,s}^0 \rangle}{\langle f_{\rm CP}|H_{eff}|B_{d,s}^0 \rangle}$ with $\eta_f$
being the {\it CP}-eigenvalue of the final states.
} in the PQCD approach
can be read as follows,
\begin{itemize}
\item[]{(a) For the $B_d^0 \to K_0^{*+} K_0^{*-}$ decays, }
\beq
{\cal A}_{\rm dir}(B_d^0 \to \kappa^+ \kappa^-) &=& 15.5^{+0.0}_{-0.6}(\omega_B)^{+3.7}_{-5.4}(B_{m})^{+0.0}_{-0.0}(\bar f_{K_0^*})^{+0.7}_{-0.8}(V)
[15.5^{+3.8}_{-5.5}]
\times 10^{-2}\;,
\label{eq:kap-d}
\eeq
\beq
{\cal A}_{\rm mix}(B_d^0 \to \kappa^+ \kappa^-) &=& -80.4^{+0.1}_{-0.2}(\omega_B)^{+3.5}_{-3.8}(B_{m})^{+0.0}_{-0.0}(\bar f_{K_0^*})^{+3.8}_{-3.3}(V)[-80.4^{+5.2}_{-5.0}]
\times 10^{-2}\;;
\label{eq:kap-m}
\eeq
and
\beq
{\cal A}_{\rm dir}(B_d^0 \to {K_0^*}(1430)^+ {K_0^*}(1430)^-) &=&
\left\{ \begin{array}{ll}
-73.9^{+0.8}_{-0.0}(\omega_B)
%^{+3.6+2.9}_{-3.9-2.6}(B_{m})^{+0.0}_{-0.2}(\bar f_{K_0^*})^{+2.5+0.3}_{-2.3-0.2}(V)
^{+4.6}_{-4.7}(B_{m})^{+0.0}_{-0.2}(\bar f_{K_0^*})^{+2.5}_{-2.3}(V)[-73.9^{+5.3}_{-5.2}]
\times 10^{-2}({\rm S1})\;,&  \\
\hspace{0.28cm}21.4^{+0.6}_{-0.8}(\omega_B)
%^{+2.6+8.4}_{-2.7-0.0}(B_{m})^{+0.0}_{-0.7}(\bar f_{K_0^*})^{+0.8+0.5}_{-0.7-0.4}(V)
^{+8.8}_{-2.7}(B_{m})^{+0.0}_{-0.7}(\bar f_{K_0^*})^{+0.9}_{-0.8}(V)[\hspace{0.28cm}21.4^{+8.9}_{-3.0}]
\times 10^{-2}({\rm S2})\;; &   \\ \end{array} \right.
\label{eq:k0s-d}
\eeq
\beq
{\cal A}_{\rm mix}(B_d^0 \to {K_0^*}(1430)^+ {K_0^*}(1430)^-) &=&
\left\{ \begin{array}{ll}
-39.6^{+5.6}_{-6.3}(\omega_B)
%^{+5.6+2.5}_{-5.1-3.0}(B_{m})^{+0.3}_{-0.0}(\bar f_{K_0^*})^{+4.1+3.9}_{-4.0-3.5}(V)
^{+6.1}_{-5.9}(B_{m})^{+0.3}_{-0.0}(\bar f_{K_0^*})^{+5.7}_{-5.3}(V)[-39.6^{+10.1}_{-10.1}]
\times 10^{-2}({\rm S1})\;,&  \\
-97.5^{+0.1}_{-0.2}(\omega_B)
%^{+0.4+2.5}_{-0.4-0.0}(B_{m})^{+0.0}_{-0.2}(\bar f_{K_0^*})^{+0.1+0.4}_{-0.2-0.3}(V)
^{+2.5}_{-0.4}(B_{m})^{+0.0}_{-0.2}(\bar f_{K_0^*})^{+0.4}_{-0.4}(V)[-97.5^{+2.5}_{-0.6}]
\times 10^{-2}({\rm S2})\;; &   \\ \end{array} \right.
\label{eq:k0s-m}
\eeq

\item[]{(b) For the $B_s^0 \to a_0^+ a_0^-$ decays, }
\beq
{\cal A}_{\rm dir}(B_s^0 \to a_0(980)^+ a_0(980)^-) &=&
-0.7^{+0.0}_{-0.0}(\omega_B)
%^{+0.1+0.5}_{-0.1-0.1}(B_{m})^{+0.0}_{-0.0}(\bar f_{a_0})^{+0.0+0.1}_{-0.0-0.0}(V)
^{+0.5}_{-0.1}(B_{m})^{+0.0}_{-0.0}(\bar f_{a_0})^{+0.1}_{-0.0}(V)[-0.7^{+0.5}_{-0.1}]
\times 10^{-2}\;,
\label{eq:a98-d}
\eeq
\beq
{\cal A}_{\rm mix}(B_s^0 \to a_0(980)^+ a_0(980)^-) &=&
13.6^{+0.0}_{-0.1}(\omega_B)
%^{+0.0+1.0}_{-0.0-1.0}(B_{m})^{+0.0}_{-0.0}(\bar f_{a_0})^{+0.0+0.5}_{-0.0-0.4}(V)
^{+1.0}_{-1.0}(B_{m})^{+0.0}_{-0.0}(\bar f_{a_0})^{+0.5}_{-0.4}(V)[13.6^{+1.1}_{-1.1}]
\times 10^{-2}\;,
\label{eq:a98-m}
\eeq
\beq
{\cal A}_{\Delta\Gamma_s}(B_s^0 \to a_0(980)^+ a_0(980)^-) &=&
99.1^{+0.0}_{-0.0}(\omega_B)
%^{+0.0+0.1}_{-0.0-0.2}(B_{m})^{+0.0}_{-0.0}(\bar f_{a_0})^{+0.0+0.0}_{-0.0-0.1}(V)
^{+0.1}_{-0.2}(B_{m})^{+0.0}_{-0.0}(\bar f_{a_0})^{+0.0}_{-0.1}(V)[99.1^{+0.1}_{-0.2}]
\times 10^{-2}\;;
\label{eq:a98-t}
\eeq
and
\beq
{\cal A}_{\rm dir}(B_s^0 \to a_0(1450)^+ a_0(1450)^-) &=&
\left\{ \begin{array}{ll}
9.6^{+0.9}_{-0.8}(\omega_B)
%^{+0.8+0.5}_{-0.9-0.8}(B_{m})^{+0.0}_{-0.1}(\bar f_{a_0})^{+0.0+0.3}_{-0.1-0.3}(V)
^{+0.9}_{-1.2}(B_{m})^{+0.0}_{-0.1}(\bar f_{a_0})^{+0.3}_{-0.3}(V)[9.6^{+1.3}_{-1.5}]
\times 10^{-2}({\rm S1})\;,&  \\
1.4^{+0.1}_{-0.0}(\omega_B)
%^{+0.3+0.2}_{-0.2-0.5}(B_{m})^{+0.1}_{-0.0}(\bar f_{a_0})^{+0.1+0.1}_{-0.0-0.0}(V)
^{+0.4}_{-0.5}(B_{m})^{+0.1}_{-0.0}(\bar f_{a_0})^{+0.1}_{-0.0}(V)[1.4^{+0.4}_{-0.5}]
\times 10^{-2}({\rm S2})\;; &   \\ \end{array} \right.
\label{eq:a14-d}
\eeq
\beq
{\cal A}_{\rm mix}(B_s^0 \to a_0(1450)^+ a_0(1450)^-) &=&
\left\{ \begin{array}{ll}
14.0^{+0.0}_{-0.1}(\omega_B)
%^{+0.9+1.0}_{-1.3-0.9}(B_{m})^{+0.0}_{-0.0}(\bar f_{a_0})^{+0.0+0.5}_{-0.0-0.4}(V)
^{+1.3}_{-1.6}(B_{m})^{+0.0}_{-0.0}(\bar f_{a_0})^{+0.5}_{-0.4}(V)[14.0^{+1.4}_{-1.7}]
\times 10^{-2}({\rm S1})\;,&  \\
\hspace{0.18cm}7.2^{+0.1}_{-0.0}(\omega_B)
%^{+0.6+0.1}_{-0.7-0.0}(B_{m})^{+0.0}_{-0.0}(\bar f_{a_0})^{+0.0+0.3}_{-0.0-0.2}(V)
^{+0.6}_{-0.7}(B_{m})^{+0.0}_{-0.0}(\bar f_{a_0})^{+0.3}_{-0.2}(V)[\hspace{0.18cm}7.2^{+0.7}_{-0.7}]
\times 10^{-2}({\rm S2})\;; &   \\ \end{array} \right.
\label{eq:a14-m}
\eeq
\beq
{\cal A}_{\Delta\Gamma_s}(B_s^0 \to a_0(1450)^+ a_0(1450)^-) &=&
\left\{ \begin{array}{ll}
98.6^{+0.0}_{-0.1}(\omega_B)
%^{+0.0+0.0}_{-0.1-0.1}(B_{m})^{+0.0}_{-0.0}(\bar f_{a_0})^{+0.0+0.0}_{-0.1-0.1}(V)
^{+0.0}_{-0.1}(B_{m})^{+0.0}_{-0.0}(\bar f_{a_0})^{+0.0}_{-0.1}(V)[98.6^{+0.0}_{-0.2}]
\times 10^{-2}({\rm S1})\;,&  \\
99.7^{+0.0}_{-0.0}(\omega_B)
%^{+0.1+0.0}_{-0.0-0.0}(B_{m})^{+0.0}_{-0.0}(\bar f_{a_0})^{+0.0+0.0}_{-0.0-0.0}(V)
^{+0.1}_{-0.0}(B_{m})^{+0.0}_{-0.0}(\bar f_{a_0})^{+0.0}_{-0.0}(V)[99.7^{+0.1}_{-0.0}]
\times 10^{-2}({\rm S2})\;, &   \\ \end{array} \right.
\label{eq:a14-t}
\eeq
\end{itemize}
in which the Gegenbauer moments in the scalar meson distribution amplitudes and the parameters
in the CKM matrix elements contribute to the majored errors theoretically, as clearly seen
from the above Eqs.~(\ref{eq:kap-d})-(\ref{eq:a14-t}).

Some comments are in order:
\begin{itemize}
\item
It is clear to see that these predicted {\it CP} violations are generally
insensitive to the variation of the scalar decay constant $\bar f_{M}(M=a_0,\; K_0^*)$.
The underlying reason is that the decay amplitudes of the considered decays are
nearly proportional to the scalar decay constant, due to the vanishing vector decay constant
(See Eqs.~(\ref{eq:decs-ka})-(\ref{eq:gb-a14}) for detail)
in the leading twist distribution amplitude~\cite{Cheng:2005nb},
\beq
\phi_{S}(x,\mu)&=&\frac{3}{\sqrt{6}}x(1-x)\biggl\{f_{S}(\mu)+\bar
f_{S}(\mu)\sum_{m=1}^\infty B_m(\mu)C^{3/2}_m(2x-1)\biggr\}\;,
\label{eq:t2-S}
\eeq
where $f_{S}(\mu)$ and $\bar f_{S}(\mu)$, $B_m(\mu)$, and
$C_m^{3/2}(t)$ are the vector and the scalar decay constants, the
Gegenbauer moments, and the Gegenbauer polynomials, respectively,
and the asymptotic forms of the twist-3
distribution amplitudes of the scalar $a_0$ and $K_0^*$ mesons.
While, the {\it CP} asymmetries of the $B_d^0 \to {K_0^*}(1430)^+ {K_0^*}(1430)^-$
decays are more sensitive to the $\bar f_{K_0^*(1430)}$ than those of the
$B_s^0 \to a_0(1450)^+ a_0(1450)^-$ ones to the $\bar f_{a_0(1450)}$. The fact is
that the isospin symmetry breaking effect from the $u$ and $d$ quark masses
leads to the tiny and negligible vector decay constant $f_{a_0(1450)}$,
i.e., Eq.~(\ref{eq:decs-a14}), and the $SU(3)$ flavor symmetry
breaking effect from the $u$ and $s$ quark masses results in the small but non-negligible
$f_{K_0^*(1430)}$, i.e., Eq.~(\ref{eq:decs-k0s}).

\item
It is easy to find that, apart from the ${\cal A}_{\rm dir}(B_s^0 \to a_0(980)^+
a_0(980)^-)_{\rm S1}$ and ${\cal A}_{\rm dir}(B_s^0 \to a_0(1450)^+ a_0(1450)^-)_{\rm S2}$
with few percent, the rest {\it CP}-violating asymmetries for the considered pure
annihilation decays of $B_d^0 \to K_0^{*+} K_0^{*-}$ and $B_s^0 \to a_0^+ a_0^-$ in the
PQCD approach are large, which means that the contributions from the penguin diagrams
are generally sizable. To see this point explicitly, we present the decay amplitudes
classified as the tree diagrams and the penguin ones, respectively, in the
Tables~\ref{tab:DeAms-tp-S1}-\ref{tab:DeAms-tp-S2}, where only the central values
are quoted for clarifications. Then it is expected that these predictions of the
{\it CP} violations, associated with
the predicted large decay rates, could be confronted with the relevant experiments at
LHCb and/or Belle-II in the (near) future. Of course, the ${\cal A}_{\rm dir}(B_s^0
\to a_0(980)^+ a_0(980)^-)_{\rm S1}$ and ${\cal A}_{\rm dir}(B_s^0 \to a_0(1450)^+
a_0(1450)^-)_{\rm S2}$ are too small to be measured easily in the near future, though the
corresponding branching ratios are as large as $10^{-5}$.
%%%================================================================================
\begin{table}[htb]
\caption{The tree and penguin decay amplitudes(in units of $10^{-3}$~GeV$^{3}$) in S1
of the pure annihilation $B_d^0 \to K_{0}^{*+} K_{0}^{*-}$ and
$B_s^0\to a_{0}^+ a_{0}^-$ decays in the PQCD approach, where the results in the parentheses
are the corresponding amplitudes of the $\bar B_d^0 \to K_{0}^{*+} K_{0}^{*-}$ and
$\bar B_s^0\to a_{0}^+ a_{0}^-$ decays, and only the central values are considered for
clarifications. }
\label{tab:DeAms-tp-S1}
\begin{center}\vspace{-0.5cm}{%\footnotesize
\begin{tabular}[t]{c||c|c}
\hline  \hline
   Modes
   & Tree diagrams(T3A) & Penguin diagrams(T3A)  \\
\hline
 $B_d^0 \to {K_{0}^{*}(800)}^+ {K_{0}^{*}(800)}^-$
     &$\begin{array}{c} 0.551 - i\; 1.920 \;
 \\ (-1.615 + i\; 1.176) \end{array}$
     &$\begin{array}{c} -1.160 - i\; 0.054 \;
 \\ (-0.779 - i\; 0.861) \end{array}$
 \\
 \hline
 $B_d^0 \to {K_{0}^{*}(1430)}^+ {K_{0}^{*}(1430)}^-$
     &$\begin{array}{c} 1.633 - i\; 1.433 \;
 \\ (-2.169 + i\; 0.126) \end{array}$
     &$\begin{array}{c} 0.245 - i\; 1.222 \;
 \\ (1.040 - i\; 0.687) \end{array}$
 \\
 \hline\hline
 $B_s^0 \to a_0(980)^+ a_0(980)^-$
     &$\begin{array}{c} -0.013 - i\; 0.504 \;
 \\ (-0.300 + i\; 0.406) \end{array}$
     &$\begin{array}{c} 6.906 + i\; 1.827 \;
 \\ (6.906 + i\; 1.827) \end{array}$
 \\
 \hline
 $B_s^0 \to a_0(1450)^+ a_0(1450)^-$
     &$\begin{array}{c} 0.738 - i\; 0.619 \;
 \\ (-0.963 + i\; 0.035) \end{array}$
     &$\begin{array}{c} -2.332 + i\; 10.678 \;
 \\ (-2.332 + i\; 10.678) \end{array}$
 \\
 \hline \hline
\end{tabular}}
\end{center}
\end{table}
%%===========================================================
%%
%%%================================================================================
\begin{table}[htb]
\caption{Similar to Table~\ref{tab:DeAms-tp-S1} but in S2 for the
$B_d^0 \to {K_0^*}(1430)^+ {K_0^*}(1430)^-$ and
$B_s^0\to a_{0}(1450)^+ a_{0}(1450)^-$ decays.
}
\label{tab:DeAms-tp-S2}
\begin{center}\vspace{-0.5cm}{%\footnotesize
\begin{tabular}[t]{c||c|c||c|c}
\hline  \hline
   Modes
   & Tree diagrams(T3A) &Penguin diagrams(T3A) & Tree diagrams(T3G) &Penguin diagrams(T3G) \\
 \hline
 $B_d^0 \to {K_{0}^{*}(1430)}^+ {K_{0}^{*}(1430)}^-$
    & $\begin{array}{c} -1.081 - i\; 1.178 \;
 \\ (0.129 + i\; 1.593) \end{array}$
    & $\begin{array}{c} -0.859 + i\; 1.365 \;
 \\ (-1.574 + i\; 0.352) \end{array}$
    & $\begin{array}{c} -0.566 + i\; 0.658 \;
 \\ (0.850 - i\; 0.171) \end{array}$
    & $\begin{array}{c} -1.079 + i\; 1.675 \;
 \\ (-1.949 + i\; 0.415) \end{array}$
 \\
 \hline\hline
 $B_s^0 \to a_0(1450)^+ a_0(1450)^-$
     & $\begin{array}{c} -0.328 - i\; 0.048 \;
 \\ (0.229 + i\; 0.239) \end{array}$
     & $\begin{array}{c} 5.364 - i\; 6.730 \;
 \\ (5.364 - i\; 6.730) \end{array}$
     & $\begin{array}{c} -0.097 + i\; 0.188 \;
 \\ (0.192 - i\; 0.088) \end{array}$
     & $\begin{array}{c} 4.954 - i\; 6.998 \;
 \\ (4.954 - i\; 6.998) \end{array}$
 \\
 \hline \hline
\end{tabular}}
\end{center}
\end{table}
%%=====================================

\end{itemize}

At last, we shall discuss the effects arising from the twist-3 distribution amplitudes
of the scalar
$a_0(1450)$ and $K_0^*(1430)$ mesons by including
the Gegenbauer polynomials in S2,
as mentioned in the above footnote~2.
It is noted that the twist-3 distribution amplitudes with Gegenbauer polynomials
of the scalar mesons in ${\rm S2}$ have been investigated in Ref.~\cite{Lu:2006fr},
\beq
\phi_{K_0^*(1430)}^S(x) &=& \frac{\bar f_{K_0^*(1430)}}{2\sqrt{2N_c}}
 \biggl\{ 1 + a_1 (2x -1) +\frac{1}{2} a_2 \biggl( 3(2x-1)^2 -1 \biggr) \biggr\}\;,
\label{eq:k0-t3-s}
\\
\phi_{K_0^*(1430)}^T(x) &=& \frac{\bar f_{K_0^*(1430)}}{2\sqrt{N_c}}
\frac{d}{dx} \biggl\{ x(1-x) \biggl[ 1 + 3 b_1 (2x - 1) +\frac{3}{2} b_2 \biggl( 5(2x -1)^2 -1 \biggr) \biggr] \biggr\} \;,
\label{eq:k0-t3-t}
\eeq
with the Gegenbauer moments
\beq
a_1 &=&  0.030 \pm 0.012 \;, \qquad a_2 = -0.18 \pm 0.15\;,\qquad
b_1 = 0.046 \pm 0.009\;, \qquad b_2 = 0.075 \pm 0.075\;,
\label{eq:k0-t3-G}
\eeq
at $\mu =1$~GeV for the $K_0^*(1430)$ state, and
\beq
\phi_{a_0(1450)}^S(x) &=&  \frac{\bar f_{a_0(1450)}}{2\sqrt{2N_c}}\biggl\{ 1+
\frac{1}{2} a_2 \biggl(3 (2 x -1)^2 -1\biggr)
+ \frac{1}{8} a_4\biggl( 35 (2 x -1)^4 - 30 (2 x-1)^2+3 \biggr)\biggr\}\;,
\label{eq:a0-t3-s}
\\
\phi_{a_0(1450)}^T(x) &=&  \frac{\bar f_{a_0(1450)}}{2\sqrt{2N_c}}
\frac{d}{dx}\biggl\{x(1-x) \biggl[ 1+ \frac{3}{2}b_2\biggl( 5(2x -1)^2-1 \biggr) + \frac{15}{8} b_4 \biggl( 21(2x -1)^4 -14(2x -1)^2 +1 \biggr)\biggr] \biggr\}\;,
\label{eq:a0-t3-t}
\eeq
with the Gegenbauer moments
\beq
a_2 &=& -0.255 \pm 0.075\;, \qquad a_4 = 0.14 \pm 0.25\;, \qquad
b_2 = 0.029 \pm 0.029\;, \qquad b_4= 0.135 \pm 0.065\;,
\label{eq:a0-t3-G}
\eeq
at $\mu =1$~GeV for the $a_0(1450)$ state, respectively~\footnote{It is necessary to
mention that the Gegenbauer moments for the twist-3 distribution amplitudes of $K_0^*(1430)$
and $a_0(1450)$ states are originally presented as~\cite{Lu:2006fr} (1) $a_1 = 0.018
\sim 0.042$, $a_2 = -0.33 \sim -0.025$, $b_1 = 0.037 \sim 0.055$, and $b_2 = 0 \sim 0.15$
at $\mu =1$~GeV for the $K_0^*(1430)$ state, and (2)  $a_2 = -0.33 \sim -0.18$, $a_4 =-0.11
\sim 0.39$, $b_2 = 0 \sim 0.058$, and $b_4= 0.070 \sim 0.20$ for the $a_0(1450)$ state,
respectively. To give the numerical results
as central values with uncertainties, we adopt the form of the Gagenbauer moments
in the twist-3 distribution amplitudes as presented in the Eqs.~(\ref{eq:k0-t3-G})
and (\ref{eq:a0-t3-G}) for convenience.}.
Then, by including the contributions from these twist-3 distribution amplitudes, the numerical
results for the branching ratios and the {\it CP}-violating asymmetries in ${\rm S2}$ could
be obtained as follows,
\beq
Br(B_d^0 \to {K_0^*}(1430)^+ {K_0^*}(1430)^-) &=&
2.38^{+0.31}_{-0.27}(\omega_B)
%^{+0.94+1.15}_{-0.53-0.69}(B_m)^{+0.00+0.07+0.03+3.37}_{-0.00-0.05-0.03-0.32}(ab)
%^{+1.27}_{-0.90}(\bar f_{K_0^*})^{+0.05+0.04}_{-0.05-0.04}(V)
^{+1.49}_{-0.87}(B_m)^{+3.37}_{-0.33}(ab)
^{+1.27}_{-0.90}(\bar f_{K_0^*})^{+0.06}_{-0.06}(V)[2.38^{+3.91}_{-1.32}]
\times 10^{-6}\;,
%\non
%&=& 2.38^{+3.91}_{-1.32}\times 10^{-6}\;,
\eeq
%%%and
\beq
Br(B_s^0 \to a_0(1450)^+ a_0(1450)^-) &=&
3.60^{+0.98}_{-0.77}(\omega_B)
%^{+2.11+0.97}_{-1.47-0.72}(B_m)^{+0.10+0.34+0.01+0.19}_{-0.10-0.32-0.02-0.13}(ab)
%^{+1.84}_{-1.33}(\bar f_{a_0})^{+0.00}_{-0.00}(V)
^{+2.32}_{-1.64}(B_m)^{+0.40}_{-0.36}(ab)
^{+1.84}_{-1.33}(\bar f_{a_0})^{+0.00}_{-0.00}(V)[3.60^{+3.14}_{-2.28}]
\times 10^{-5}\;,
%\non
%&=&3.60^{+3.14}_{-2.28} \times 10^{-5}\;,
\eeq
in which the central value of the branching fraction $Br(B_d^0 \to {K_0^*}(1430)^+
{K_0^*}(1430)^-)_{\rm S2}[Br(B_s^0 \to a_0(1450)^+ a_0(1450)^-)_{\rm S2}]$ becomes
slightly smaller[larger] than that obtained in the case with adopting the asymptotic
form of the twist-3 distribution amplitudes correspondingly.
Of course, the {\it CP}-averaged branching ratios almost remain unchanged within the still large
theoretical errors. And
\beq
{\cal A}_{\rm dir}(B_d^0 \to {K_0^*}(1430)^+ {K_0^*}(1430)^-)&=&
-73.1^{+11.0}_{-8.1}(\omega_B)
%^{+19.7+49.1}_{-11.7-5.8}(B_m)^{+0.5+4.9+0.3+89.1}_{-0.5-4.5-0.3-10.2}(ab)
%^{+0.0}_{-0.3}(\bar f_{K_0^*})^{+1.6+1.2}_{-1.7-1.2}(V)
^{+52.9}_{-13.1}(B_m)^{+89.2}_{-11.2}(ab)
^{+0.0}_{-0.3}(\bar f_{K_0^*})^{+2.0}_{-2.1}(V)[-73.1^{+104.3}_{-19.2}]
\times 10^{-2}\;,
\label{eq:br-t3g}
%\non
%&=& -73.1^{+104.3}_{-19.2} \times 10^{-2}\;,
\\
{\cal A}_{\rm mix}(B_d^0 \to {K_0^*}(1430)^+ {K_0^*}(1430)^-)&=&
\hspace{0.18cm}-3.5^{+0.7}_{-0.0}(\omega_B)
%^{+33.6+48.5}_{-32.4-38.2}(B_m)^{+0.8+8.1+2.1+56.6}_{-0.7-7.1-2.0-94.8}(ab)
%^{+0.3}_{-0.0}(\bar f_{K_0^*})^{+1.2+0.8}_{-1.2-0.6}(V)
^{+59.0}_{-50.1}(B_m)^{+57.2}_{-95.1}(ab)
^{+0.3}_{-0.0}(\bar f_{K_0^*})^{+1.4}_{-1.3}(V)[-3.5^{+82.2}_{-107.5}]
\times 10^{-2}\;,
%\non
%&=&-3.5^{+82.2}_{-107.5} \times 10^{-5}\;;
\eeq
%%and
\beq
{\cal A}_{\rm dir}(B_s^0 \to a_0(1450)^+ a_0(1450)^-)&=&
\hspace{0.18cm}4.6^{+0.4}_{-0.5}(\omega_B)
%^{+1.9+0.9}_{-1.3-1.9}(B_m)^{+0.3+0.4+0.4+0.9}_{-0.3-0.4-0.4-1.1}(ab)
%^{+0.0}_{-0.0}(\bar f_{a_0})^{+0.1}_{-0.2}(V)
^{+2.1}_{-2.3}(B_m)^{+1.1}_{-1.3}(ab)
^{+0.0}_{-0.0}(\bar f_{a_0})^{+0.1}_{-0.2}(V)[4.6^{+2.6}_{-2.7}]
\times 10^{-2}\;,
%\non
%&=&4.6^{+2.6}_{-2.7} \times 10^{-2}\;,
\\
{\cal A}_{\rm mix}(B_s^0 \to a_0(1450)^+ a_0(1450)^-)&=&
\hspace{0.18cm}8.6^{+0.7}_{-0.3}(\omega_B)
%^{+17.0+15.9}_{-23.0-14.2}(B_m)^{+1.9+6.5+3.5+27.3}_{-2.0-7.1-3.7-25.3}(ab)
%^{+0.1}_{-0.0}(\bar f_{a_0})^{+0.3}_{-0.3}(V)
^{+23.3}_{-27.0}(B_m)^{+28.3}_{-26.6}(ab)
^{+0.1}_{-0.0}(\bar f_{a_0})^{+0.3}_{-0.3}(V)[8.6^{+36.7}_{-37.9}]
\times 10^{-3}\;,
%\non
%&=&8.6^{+36.7}_{-37.9}\times 10^{-3}\;,
\\
{\cal A}_{\rm \Delta\Gamma_s}(B_s^0 \to a_0(1450)^+ a_0(1450)^-)&=&
99.9^{+0.0}_{-0.0}(\omega_B)^{+0.1}_{-0.1}(B_m)^{+0.0}_{-0.1}(ab)
^{+0.0}_{-0.0}(\bar f_{a_0})^{+0.0}_{-0.0}(V)[99.9^{+0.1}_{-0.1}]
\times 10^{-2}\;,
\label{eq:3cp-t3g}
%\non
%&=&99.9^{+0.1}_{-0.1} \times 10^{-2}\;.
\eeq
where the direct {\it CP}-violating asymmetry and the mixing-induced one
are changed significantly for the $B_d^0 \to {K_0^*}(1430)^+ \\ {K_0^*}(1430)^-$ and
$B_s^0 \to a_0(1450)^+ a_0(1450)^-$ decays in S2. Note that the parameter $(ab)$ in
the above Eqs.~(\ref{eq:br-t3g})-(\ref{eq:3cp-t3g}) denotes the combined
Gegenbauer moment of $a_i$ and $b_i$
in the twist-3 distribution amplitudes with Gegenbauer polynomials.
In terms of the central value, the ${\cal A}_{\rm dir}$ varies from $21.4\%$ in the
T3A form to $-73.1\%$ in the T3G form for the former mode, while from $1.4\%$ in the
T3A form to $4.6\%$ in the T3G form for the latter one;
and the ${\cal A}_{\rm mix}$ changes from $-97.5\%$ to $-3.5\%$ for the former
channel, while from $7.2\%$ to $0.86\%$ for the latter one, which indicate that
the strong phases
in these two pure annihilation decays could be dramatically affected by the twist-3
distribution amplitudes with inclusion of the Gegenbauer polynomials.
In order to show the significant changes of the mentioned strong phases,
we present the decay amplitudes of these two pure annihilation modes, i.e., $B_d^0
\to {K_0^*}(1430)^+{K_0^*}(1430)^-$ and $B_s^0 \to a_0(1450)^+ a_0(1450)^-$,
in S2 explicitly by considering the twist-3 distribution amplitudes in the T3A
and T3G forms respectively, as given in the Tables~\ref{tab:DeAms-S2}
and~\ref{tab:DeAms-tp-S2}. It is clearly seen that the magnitudes of the
dominant contributions from the nonfactorizable annihilation diagrams vary
slightly, but the strong phases in both of the factorizable and
nonfactorizable annihilation diagrams change remarkably, which means that
the considered T3G contributions are also important to the strong phases in
the annihilation diagrams of this work. Therefore,
the clear understanding of the hadron dynamics about
the scalar mesons considered in this work
could be very helpful to provide precise predictions
theoretically, even to explore the presently
unknown annihilation decay mechanism.

\bigskip

In short, it is well known that the annihilation diagrams, although which
are power suppressed and with currently unknown mechanism,
could play important roles in the heavy flavor $B$ and $D$
meson decays. As one of the popular tools theoretically, the PQCD approach
has the advantages in computing the annihilation contributions. Therefore,
by assuming the scalar $a_0$ and $K_0^*$ being the
$q\bar q$ mesons in two different scenarios, we have investigated
the {\it CP}-averaged branching ratios and the
{\it CP}-violating asymmetries for the pure annihilation decays of
$B_d^0 \to K_0^{*+} K_0^{*-}$ and $B_s^0 \to a_0^+ a_0^-$ in
the PQCD approach, which embrace the same quark structure as $B_d^0 \to K^+ K^-$
and $B_s^0 \to \pi^+ \pi^-$ in the pseudoscalar sector. At the same time,
for the $B_d^0 \to {K_0^*}(1430)^+
{K_0^*}(1430)^-$ and $B_s^0 \to a_0(1450)^+ a_0(1450)^-$ decays, we also studied
the contributions from the twist-3 distribution amplitudes in the Gegenbauer polynomial
forms of the $K_0^*(1430)$ and $a_0(1450)$ in S2.
Based on the numerical results within still large theoretical errors and the
phenomenological analyses, the predictions in the PQCD approach showed that:
\begin{itemize}
\item
the large decay rates in the order of $10^{-6}$ and $10^{-5}$ have been obtained
in the PQCD calculations for the pure annihilation $B_d^0 \to K_0^{*+} K_0^{*-}$ and
$B_s^0 \to a_0^+ a_0^-$ decays, respectively, but which suffer from large theoretical
uncertainties mainly arising from the hadronic parameters of the scalar $a_0$
and $K_0^*$ mesons, such as the Gegenbauer moments, the scalar decay constants, etc.
Undoubtedly, these large predictions in the PQCD approach could be examined
at the LHCb and/or Belle-II experiments in the
(near) future.

\item
except for the ${\cal A}_{\rm dir}(B_s^0 \to a_0(980)^+
a_0(980)^-)_{\rm S1}$ and ${\cal A}_{\rm dir}(B_s^0 \to a_0(1450)^+ a_0(1450)^-)_{\rm S2}$
in the order of $1\%$, the large {\it CP}-violating asymmetries
could
be found in the rest $B_d^0 \to K_0^{*+} K_0^{*-}$ and $B_s^0 \to a_0^+ a_0^-$ modes. These
asymmetries could also be tested in the future measurements with the help of the predicted
large branching ratios.

\item
the $B_d^0 \to {K_0^*}(1430)^+
{K_0^*}(1430)^-$ and $B_s^0 \to a_0(1450)^+ a_0(1450)^-$ decays are investigated in S2
with two different kinds of twist-3 distribution amplitudes, i.e.,
the asymptotic form and
the Gegenbauer polynomial form. And the latter form could change
the strong phases evidently, and subsequently change the {\it CP}-violating asymmetries apparently.

\item
the unknown inner structure or QCD dynamics of the scalar states demands the reliable
studies from the experimental examinations, or in the nonperturbative techniques, for example, Lattice QCD. The clear understanding about the QCD dynamics of the scalars
must constrain the errors of the predictions in the PQCD approach of the pure annihilation
modes in this work, which would provide opportunities to shed light on the mechanism
of the annihilation decays.

\end{itemize}

\begin{acknowledgments}
Y.~Chen thanks L.~Su for helpful discussions.
This work is supported in part by the National Natural Science
Foundation of China under Grant Nos.~11765012 and 11205072,
and by the Research
Fund of Jiangsu Normal University(No.~HB2016004). Y.C. is supported by the
Undergraduate Research $\&$ Practice Innovation Program
of Jiangsu Province(No.~201810320103Z).
\end{acknowledgments}

%%====================================================================
%%%%%%%%%%%%%%%%%%%%    References  ==================================
%%%%==================================================================

\end{CJK*}
\end{document}